\documentclass{PoS}

\usepackage{amssymb}
\usepackage[intlimits]{amsmath}
\usepackage{amsfonts}
\usepackage{dsfont}
\usepackage{multirow}
\usepackage{subfigure}
\newcommand{\sh}[1]{#1\hskip-7pt \diagup}

\def\i{\ensuremath{\mathrm{i}}}

\def\T{\ensuremath{\mathrm{T}}}

\title{The role of vector meson dominance and running masses 
in the hadronic contributions to the muon $g-2$}

\ShortTitle{Role of VMD in hadronic LBL contribution to the muon $g-2$}

\author{\speaker{Tobias Goecke}\\
        Institut f\"ur Theoretische Physik\\
	Universit\"at Giessen, Heinrich-Buff-Ring 16, 35392 Giessen, Germany\\ 
        E-mail: \email{Tobias.Goecke@theo.physik.uni-giessen.de}}

	\author{\speaker{Christian S. Fischer}\\
Institut f\"ur Theoretische Physik\\
Universit\"at Giessen, Heinrich-Buff-Ring 16, 35392 Giessen, Germany}

\author{Richard Williams\\
Institut f\"ur Physik\\
Karl-Franzens--Universit\"at Graz, Universit\"atsplatz 5, 8010 Graz, Austria		
}

\abstract{We summarize our recent results for the quark loop part of the  light-by-light scattering
	contribution as well as the hadronic vacuum polarisation contributions
	to the anomalous magnetic moment of the muon. In particular
	we focus on the role played by the momentum dependence of the quark- and 
	quark-photon vertex dressing functions. We give a detailed comparison of the Dyson-Schwinger
	description of this contribution to the corresponding picture emerging from hadronic models
	in particular the extended Nambu--Jona-Lasonio model (ENJL). We find that the details of the
	momentum dependence are important on a quantitative level. Especially the transverse parts of 
	the quark-photon-vertex, which serve as a dynamical extension of simple vector meson dominance models,
	do not yield the large suppression of the light-by-light contribution found in the ENJL model if 
	realistic dressings are taken into account.}

\FullConference{Xth Quark Confinement and the Hadron Spectrum,\\
		October 8-12, 2012\\
		TUM Campus Garching, Munich, Germany}

\begin{document}

\section{Introduction}

\begin{table}[b]
  \centering

  \begin{tabular}{c||r|l|l}
    Contribution & $\hspace{-0.6cm}a_\mu\times 10^{11}\hspace{0.5cm}$ & 
    $\hspace{0.5cm}\dfrac{a_\mu^i}{a_\mu^{SM}}$     & $\left(\dfrac{\delta a_\mu^i}{\delta a_\mu^{SM}}\right)^2$  \\
    \hline\hline
    QED		&	$116\,584\,718.1\,(\,\,\,0.2)$ & $99.99390\%$ & $00.00098\%$   \\
    \hline
    weak	&	$153.2\,(\,\,\,1.8)$ & $00.00013\%$ & $00.07910\%$ \\
    \hline
    QCD LOHVP	&	$6\,949.1\,(58.2)$ & $00.00596\%$ & $82.69628\%$ \\
    \hline
    QCD HOHVP	& 	$-98.4\,(\,\,\,1.0)$ & $00.00008\%$ & $00.02441\%$	\\
    \hline
    QCD LBL	&	$105\,\,\,\,\,\,(26\,\,\,\,)$ & $00.00009\%$ & $16.50391\%$ \\
    \hline
    Standard Model &   $116\,591\,827.0\,(64\,\,\,\,)$ & $100\%$ & $100\%$ \\
    \hline
    Experiment	&	$116\,592\,089\,\,\,\,\,\,(63\,\,\,\,)$ &\\
    \hline\hline
    Exp-Theo	&	$262\,\,\,\,\,\,(89\,\,\,\,)$ &
  \end{tabular}
    \caption{Standard Model contributions to the muon $g-2$.}
  \label{tab:DiffContrToAm}
\end{table}
In these proceedings we summarize our results on the anomalous magnetic moment of
the muon presented in~\cite{Goecke:2012qm} and ~\cite{Goecke:2011pe}. This quantity provides a precision 
test of the Standard Model (SM), the electromagnetic (EM)~\cite{Aoyama:2012wk},
the weak~\cite{Czarnecki:2002nt} and the strong force. An overview of the different 
contributions is given in Table~\ref{tab:DiffContrToAm}. Clearly the QED contributions 
are dominant, followed by the ones from QCD. It also evident that the strong
contributions dominate the uncertainty of the SM prediction. Most relevant
in that respect is the leading order hadronic vacuum polarisation (LOHVP)
\cite{Hagiwara:2011af}. There is some hope that this uncertainty may decrease due to 
improved experimental input~\cite{Jegerlehner:2009ry} as well as the efforts 
of the lattice community~\cite{Feng:2011zk,Boyle:2011hu,DellaMorte:2011aa}. 
In the long run the most problematic contribution may therefore be hadronic light-by-light
scattering (LBL), which cannot thus far be determined in a model-independent
way. It has been considered in many approaches such as the 
Extended Nambu--Jona-Lasinio (ENJL) model~\cite{Bijnens:1995xf}, the Hidden Local 
Symmetry (HLS) model~\cite{Hayakawa:1995ps}, vector meson dominance (VMD) 
approaches~\cite{Knecht:2001qf,Melnikov:2003xd}, the non-local chiral quark 
model (NL$\chi$QM)~\cite{Dorokhov:2008pw,Dorokhov:2012qa}, the chiral constituent 
quark model ($\chi$CQM)~\cite{Greynat:2012ww}, in holographic models~\cite{Cappiello:2010uy} 
and Dyson-Schwinger Equations (DSEs)~\cite{Fischer:2010iz,Goecke:2010if,Goecke:2012qm}.
The lattice calculations of LBL are still at an exploratory stage \cite{Blum:2013}.
The combined theory result quoted in Table~\ref{tab:DiffContrToAm} is a number that
different groups agreed upon~\cite{Prades:2009tw} by combining several hadronic models. 
For comparison, we list here the most recent experimental result from Refs.~\cite{Bennett:2006fi,Roberts:2010cj} which reveals a discrepancy of about three standard 
deviations~\cite{Hagiwara:2011af};
very recent results even indicate a significance ranging between 4.7 and 4.9 $\sigma$~\cite{Benayoun:2012wc}.

These results are exciting since they may provide a signal of new physics beyond the Standard Model. 
However, as we will argue in detail in the course of this work they should be taken with some caution. The present estimate of the hadronic light-by-light contribution
including its potential error has been derived from model calculations. Although much
care has been invested in the error estimate, we believe that the inherent limitations
of the models employed may have led to an overly optimistic value. 
Here we summarise our comparison between our approach and that of models for one of the 
more contentious contributions to LBL~\cite{Goecke:2012qm}, in order to highlight this point.
To this end we discuss the hadronic photon four-point function and summarize the DSE approach used in this work in Sections~\ref{sec:basics} and \ref{sec:method}. Then we shortly discuss the results for the leading order hadronic contribution 
in Section~\ref{sec:HVP}. In Section~\ref{sec:DSEvsENJL} we compare
our method to the extended Nambu--Jona-Lasinio model. We explain similarities and differences 
and how these affect the results for the quark-loop light-by-light contribution 
in Section~\ref{sec:results}. We conclude 
in Section~\ref{sec:conclusions}.

\begin{figure}[t!]
		\centering	\subfigure[][]{\label{fig:hadroniclo}\includegraphics[width=0.20\columnwidth]{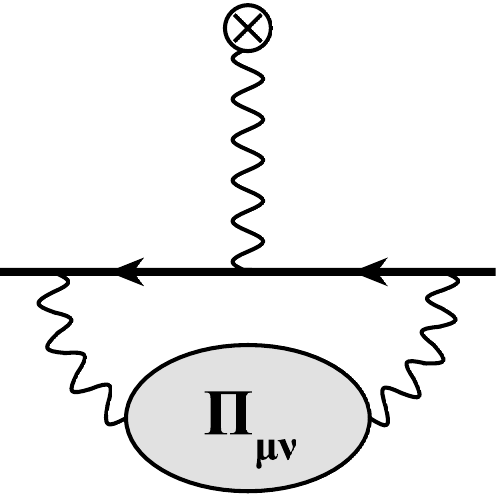}}
	\hspace{0.08\columnwidth}
	\subfigure[][]{\label{fig:hadroniclbl}\includegraphics[width=0.20\columnwidth]{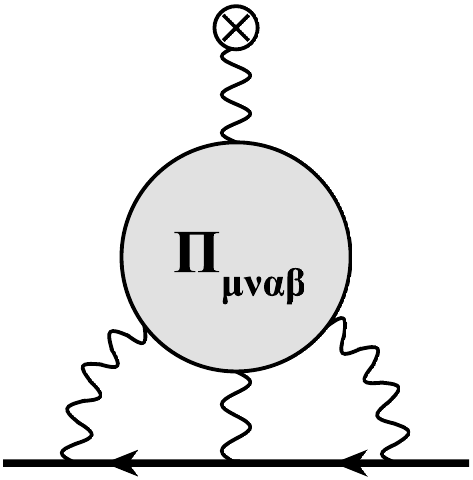}}
      \caption{The two classifications of corrections to the photon-muon
	vertex function: (a) hadronic vacuum polarization contribution to $a_\mu$. The vertex is 
               dressed by the vacuum polarization tensor $\Pi_{\mu\nu}$;
		   (b) the hadronic light-by-light scattering contribution to
		   $a_\mu$.}
\end{figure}

\section{Basics}\label{sec:basics}

On mass-shell, the muon-photon vertex can be written in terms of two momentum dependent form-factors
\begin{align}
   \parbox{1cm}{\includegraphics[width=0.08\textwidth]{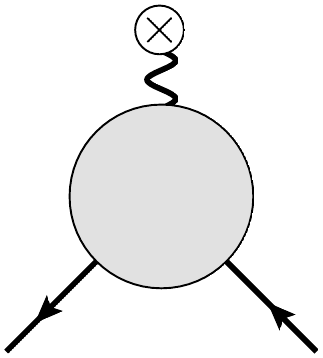}}\quad
   &=\bar{u}(p^\prime)
   \left[F_1(q^2)\gamma_\alpha+iF_2(q^2)\frac{\sigma_{\alpha\beta}q^\beta}{2 m_\mu}\right]u(p)\,,
  \label{eqn:MuonPhotonVertexDecomposition}
\end{align}
where $p$ and $p^\prime$ are the muon momenta, $q$ is the photon
momentum and $\sigma_{\alpha\beta}=\frac{i}{2}[\gamma_\alpha,\gamma_\beta]$. The 
anomalous magnetic moment is defined in the limit of vanishing photon momentum, $q^2$, as
\begin{align}
  a_\mu = \frac{g-2}{2}=F_2(0)\,.
  \label{eqn:DefOfAnomaly}
\end{align}
The HVP contribution, shown in Fig.~\ref{fig:hadroniclo} is defined within the DSE approach as the hadronic
photon self energy
\begin{align}
  \Pi_{\mu\nu}= \parbox{2.5cm}{\includegraphics[width=2.5cm]{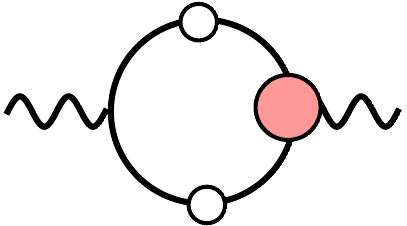}},
  \label{eqn:HadrTwoPointTenso}
\end{align}
which is itself defined in terms of the full quark and the quark-photon vertex.
This representation is exact, up to truncations of vertex and propagator.
This quantity is well described within our approach as is 
detailed in section~\ref{sec:HVP}.\\

The LBL contribution 
to this vertex is depicted in Fig.~\ref{fig:hadroniclbl}.
This hadronic photon four-point function can be split into several 
parts, which we organize by the number of quark T-Matrices involved
\begin{align}
\Pi_{\mu\nu\alpha\beta}=
  \begin{array}{c}
        \includegraphics[width=0.6\columnwidth]{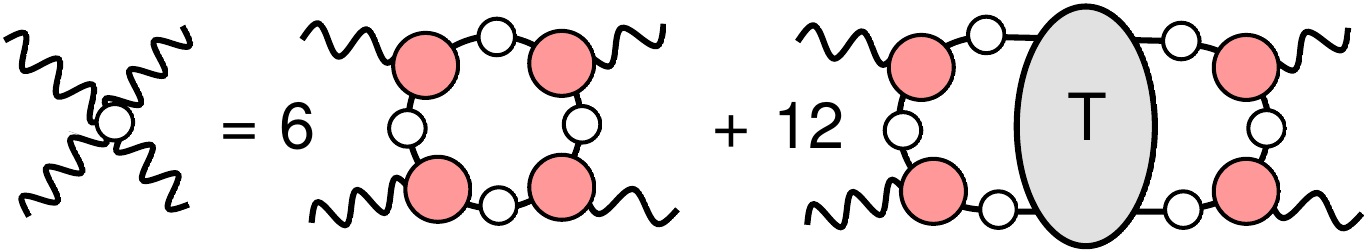}
  \end{array}+\cdots\,.
  \label{eqn:Hadr4PointFunctionDecomposition}
\end{align}
The pre factors indicate the number of permutations that must be computed.
The second term, containing just one T-matrix, has been considered in Refs.~\cite{Fischer:2010iz,Goecke:2010if} by performing a pseudoscalar resonant expansion.
We found good agreement with corresponding results from hadronic models
\cite{Bijnens:1995xf,Bijnens:2001cq,Hayakawa:1996ki,Hayakawa:1997rq,Hayakawa:2001bb,
Knecht:2001qf,Melnikov:2003xd,Nyffeler:2009uw,Prades:2009tw,Dorokhov:2008pw,Dorokhov:2011zf,Greynat:2012ww}.
The ellipsis contains, for example, contributions that can be written as charged pion loops
including also the pion polarizabilities. While in earlier works these contributions
have been argued to be small, a recent work suggests substantial contributions to LBL
which need to be evaluated carefully in the future~\cite{Engel:2012xb}.
Note, however, that within the approximation of QCD, we are working with (see Sec.~\ref{sec:method})
the representation in Eq.~(\ref{eqn:Hadr4PointFunctionDecomposition}) is
consistent with the exact representation Eq.~(\ref{eqn:HadrTwoPointTenso}), a fact
that is interesting when it comes to considerations about systematic uncertainties.

\begin{figure}[t]
  \begin{center}
     \includegraphics[width=0.2\textwidth]{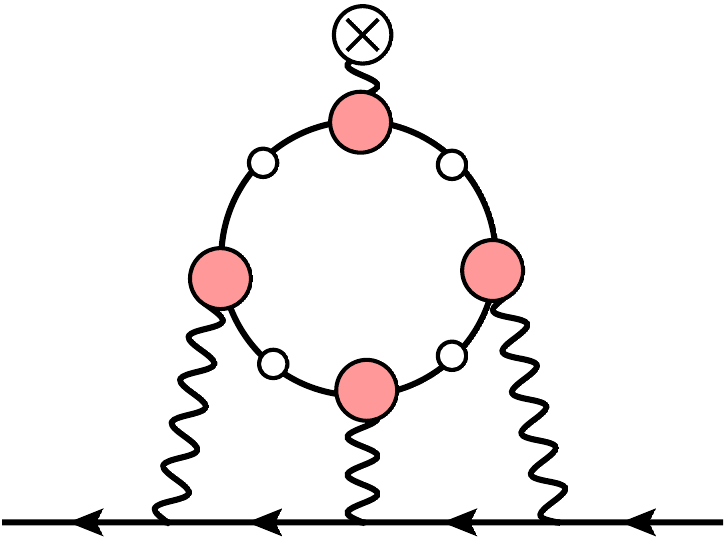} 
  \end{center}
  \caption{The quark loop contribution to the muon $g-2$. The quarks
  and vertices are fully dressed.}
  \label{fig:LBLQLcontribution}
\end{figure}

Later we concentrate on the first part of Eq.~(\ref{eqn:Hadr4PointFunctionDecomposition})
referred to here as the quark-loop. It consists of four fully dressed quark propagators as well as four
fully dressed quark-photon vertices. Its contribution to $a_\mu$ is shown explicitly in Fig.~\ref{fig:LBLQLcontribution}.
The focus of these proceedings will be in the transverse parts of these fully-dressed vertices, which in principle contain contributions from vector mesons. It has been shown that these contributions generate a strong suppression of the quark-loop contribution to $a_\mu$
in the ENJL model~\cite{Bijnens:1995xf}.
Below we will argue that this suppression is in fact a model artefact attributable to the contact interaction featured therein.

\section{Method}\label{sec:method}
First, let us briefly review the needed Dyson-Schwinger equations
and their truncation. The first ingredient for
the calculation of the quark-loop is the quark propagator, $S(p)$, whose 
inverse is described by the DSE
\begin{equation}
	S^{-1}(p)=Z_2 S_0^{-1} + g^2 Z_{1f} \frac{4}{3}\!\int\! 
	\overline{dk} \gamma^\mu S(k) \Gamma^\nu(k,p) D_{\mu\nu}(q)\,,
	\label{eqn:quark_DSE}
\end{equation}
To solve this equation we require the dressed gluon propagator 
$D_{\mu\nu}(p)$ and the dressed quark-gluon vertex $\Gamma_\nu(k,p)$.
In the rainbow truncation employed here we use a
combined ansatz for both of these quantities retaining
only the $\gamma_\nu$ component of the vertex.
This truncation has been proposed in Refs.~\cite{Maris:1997tm,Maris:1999nt}
and has achieved phenomenological success in describing meson properties
such as masses, decay constants and electromagnetic
form factors~\cite{Maris:1997tm,Maris:1999nt,Maris:1999bh,Jarecke:2002xd,Maris:2002mz}
as well as various baryon properties~\cite{Eichmann:2009qa,Eichmann:2011vu,Eichmann:2011pv,SanchisAlepuz:2011jn}.

The second ingredient for the calculation of the quark-loop is 
quark-photon vertex, $\Gamma_\mu$, determined from the inhomogeneous equation 
\begin{align}
  [\Gamma_\mu(P,k)]_{rs}=& \,\,Z_1 \gamma_\mu  
  \,\,-\,\, Z_2^2\frac{4}{3}\int \overline{dq}\,[S(q_+)\Gamma_\mu(P,q) S(q_-)]_{ut}K_{tu,rs}(k-q) \,,
\label{eqn:LadderQEDVertexBSE}
\end{align}
where $K_{tu,rs}(k-q)$ is the quark-antiquark interaction-kernel. To satisfy the
vector- and axial-vector Ward-Takahashi identities we choose the ladder-exchange kernel,
consistent with rainbow truncation of the quark DSE above. An important consequence of
this truncation is the dynamical generation of vector meson poles in the transverse 
components. Thus the tenets of vector meson dominance models (VMD) are automatically encoded
on the level of interacting quarks and gluons. Note furthermore that the vertex from
Eq.~(\ref{eqn:LadderQEDVertexBSE}) and the quark from Eq.~(\ref{eqn:quark_DSE}) fulfill
the Ward-Takahashi identity (WTI)
%
	\begin{align}
	  \i P_\mu \Gamma_\mu(P,k) &= S^{-1}(k_-) - S^{-1}(k_+).
	  \label{eqn:QEDWTI}
	\end{align}
Both properties are crucial for the realistic description of electromagnetic properties,
e.g. the pion charge radius~\cite{Maris:1999bh}. Their relevance to the leading order
hadronic vacuum polarisation contribution to $a_\mu$ has been highlighted in 
Ref.~\cite{Goecke:2011pe}, see Sec.~\ref{sec:HVP}.
We will see below also the importance of
the correct description of the transverse structures as regards the hadronic LBL 
contribution. From this we will argue that the twenty percent error quoted in Table~\ref{tab:DiffContrToAm} is too optimistic since it is obtained from approximations that
implicitly miss information that we later find to be important.

\section{The leading order QCD contribution}
\label{sec:HVP}
In the following we summarize our results for the leading order
hadronic contribution, the leading order hadronic vacuum polarisation published
in \cite{Goecke:2011pe}.
This quantity is a very important test for any approach that aims
at a quantitative prediction of the much more complicated LBL. In short,
using the quark [Eq. (\ref{eqn:quark_DSE})] and the quark-photon
vertex [Eq. (\ref{eqn:LadderQEDVertexBSE})] we calculate this contribution
via Eq. (\ref{eqn:HadrTwoPointTenso}). Our result is

$$a_\mu^{LOHVP,DSE}=(6760-7440)\times 10^{-11},$$
where the given range reflects our systematic model uncertainty, see \cite{Goecke:2011pe}
for details. Our result is
close to the rather model independent analysis from Ref.~\cite{Hagiwara:2011af}
, i.e. $6949\times 10^{-11}$. This is a deviation of less then ten percent. 
Although this precision is not a challenge for the model-independent HVP predictions,
a corresponding precision for LBL would be a great breakthrough.
Furthermore, in Ref. \cite{Goecke:2011pe} we presented a determination of the Adler function,
that can be obtained from the hadronic vacuum polarisation tensor, c.f. Eq. (\ref{eqn:HadrTwoPointTenso}),
that shows good agreement with model-independent results from dispersion relations on all scales.
What is especially interesting in the context of these proceedings is that we found
the transverse part of the quark-photon vertex to yield the dominant contribution
of $\sim 80\,\%$ to $a_\mu^{LOHVP}$. Since this part contains a dynamically generated vector meson bound
state, this explains why VMD approximations work reasonably well for HVP, see e.g. \cite{Bell:1996md}.
Models that miss these physics have to use a constituent quark mass much lower than
expected \cite{Greynat:2012ww,Boughezal:2011vw}. \\

\begin{figure}[t]
  \begin{center}
    \includegraphics[width=0.8\textwidth]{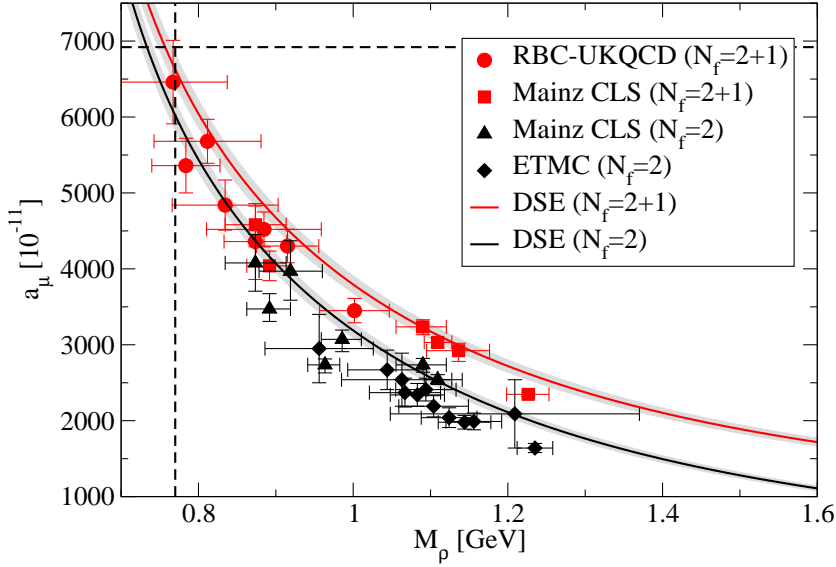}
  \end{center}
  \caption{The LOVHP contribution to $a_\mu$ as a function of the
  $\rho$ mass. We compare our results for the $N_f=2$ (black) and $N_f=2+1$ (red) flavour cases
  to lattice results. The grey bands represent our numerical uncertainty.}
  \label{fig:HVPLattice}
\end{figure}

In Ref.~\cite{Goecke:2011pe} we also compared our results against lattice QCD. 
Here, we present an updated comparison including more recent lattice data in Fig.~\ref{fig:HVPLattice}.
Both, the lattice and the DSE calculations are performed at various current quark masses.
In order to provide a gauge and scheme invariant comparison between the approaches we
have plotted the results against the corresponding varying mass of the $\rho$-meson. 
In the figure, we compare against results from Refs.~\cite{Feng:2011zk,Boyle:2011hu,DellaMorte:2010sw}
for the cases  of $N_f=2$ (black) and $N_f=2+1$ (red) flavours.
We find agreement within error bars,
showing nicely that the DSEs have the correct physics included and that
the $\rho$ meson mass is the important scale here.\\

It is interesting, that while we confirm the simple
picture of VMD for the case of the leading hadronic contribution,
the picture will be very different in the case of the LBL contribution. 
This is the subject in the following.

\section{Comparison: DSE vs. ENJL}
\label{sec:DSEvsENJL}
To explicate the differences between our approach and others, we make a comparison
between the DSE truncation used in this work and the ENJL model.
The first striking difference lies in the fact that DSEs have a smooth ultraviolet
limit, wherein at large momenta we connect with perturbation theory. ENJL on the other
hand is non-renormalisable and features an effective cut-off scale on the order of a GeV.
This necessitates the splitting of the photon four-point function into a high- and low-energy
region that is not unique due to their being two independent scales.
Additionally, the dressings of the ENJL model feature trivial momentum dependencies due
to its contact interaction. In this regard the DSE approach is quite different. In what
follows we will explore this is some detail; further information can be found in
Ref.~\cite{Goecke:2012qm}.

Primarily we will focus on the two flavour case as this is sufficient to make our point. However,
we will quote $N_f=4$ results for completeness in the later sections.

\subsection{The ENJL perspective}
The discussion here is based on Refs.~\cite{Bijnens:1995xf,Bijnens:1994ey}, 
where further details may be found. In the ENJL model, the
quark propagator assumes
the form of a momentum independent constituent quark
\begin{align}
  S^{-1}_{\mathrm{ENJL}}(p) = -i\, \sh{p}+M\,,
  \label{eqn:NJLQuark}
\end{align}
with a typical mass of $M\approx 300 \,$MeV. The
wave function renormalisation, $Z_f$, is unity.
The quark-photon vertex is given by the bubble sum
\begin{align}
  \begin{array}{c}
  \includegraphics[width=0.5\textwidth]{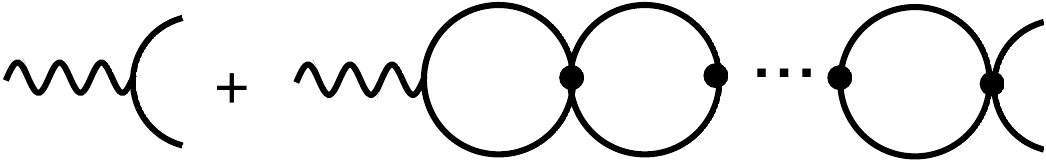}
  \end{array}
  \, , \label{eqn:NJLBubbleSum}
\end{align}
which can be resummed as a geometric series 
$\sum_n \mathrm{Bubble}^n = 1/(1-\mathrm{Bubble})$.
This simple behaviour is due to the effective contact
interaction that decouples the loop integrals. Consequently, 
the result depends only on the total photon momentum $Q$ and has
no dependence on the relative quark momentum.
Explicitly the vertex has the form
\begin{align}
  \Gamma_\mu^\mathrm{ENJL} = \gamma_\mu - \gamma^T_\mu \frac{Q^2}{Q^2+M_V^2}\,,
  \label{eqn:NJLVertexLT}
\end{align}
which contains the bare vertex $\gamma_\mu$ and the leading transverse structure
$\gamma_\mu^\T=(\delta_{\mu\nu}-Q_\mu Q_\nu/Q^2)\gamma_\nu$. The dressing of this transverse part is given here 
in the VMD limit of the ENJL model \cite{Bijnens:1994ey} where
for two flavours $M_V$ is identified with the $\rho$-mass. Using the transversality of the hadronic photon four-point
function with respect to its photon legs, this vertex can be reduced to
$\gamma_\mu M_V^2/(Q^2+M_V^2)$ which highlights the strong suppression induced by the
VMD contribution to the vertex. Note that this model is
consistent with the Ward-Takahashi identity, with the $\gamma_\mu$ component the constrained
gauge-part and $\gamma^T_\mu$ the transverse part.

\subsection{The DSE perspective}

Now we investigate the corresponding objects in the case of DSEs.
The full inverse quark propagator is given by
\begin{align}
  S^{-1}_{\mathrm{DSE}}(p) =  Z_f^{-1}(p^2)\,(-i\, \sh{p}+M(p^2))\,,
  \label{eqn:DSEQuarkProp}
\end{align}
which contains two momentum dependent dressing functions
$Z_f(p^2)$ and $M(p^2)$, which correspond to the wavefunction
renormalisation and the mass function, respectively.
The quark-photon vertex is given as a sum of gluon ladders
\begin{align}
  \begin{array}{c}
  \includegraphics[width=0.5\columnwidth]{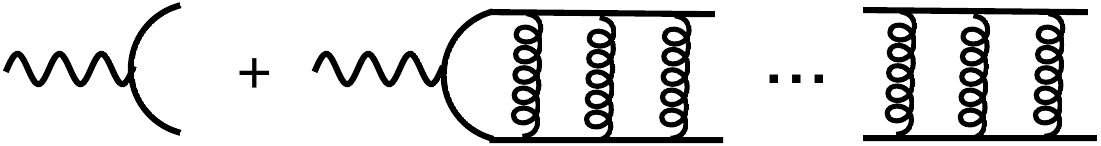}
  \end{array}\,,
  \label{eqn:DSELAdderSum}
\end{align}
which, as opposed to Eq. (\ref{eqn:NJLVertexLT}), can not
be resummed as a trivial series and must instead be solved numerically.
In general, this vertex can be decomposed into twelve tensor components
\begin{align}
  \Gamma_\mu(Q,k) = \sum_{i=1}^{4} \lambda^{(i)}(Q,k)L_\mu^{(i)} + \sum_{i=1}^{8} \tau^{(i)}(Q,k)T_\mu^{(i)} \,,
  \label{eqn:VertexDecomposition}
\end{align}
where $k$ is the relative quark momentum and $Q$ the total photon momentum.
Note that the
first part, containing the $\lambda^{(i)}$, is also
called the Ball-Chiu vertex $\Gamma_\mu^{BC}$ \cite{Ball:1980ay}. 
The vector meson bound state does also appear in the transverse
vertex structure in the DSE/BSE case. This is most easily
pictured in the form of a simple fit \cite{Maris:1999bh}
to the numerical results of the quark-photon vertex
\begin{align}
  \Gamma_\mu(Q,k) \simeq\Gamma_\mu^\mathrm{BC}\!-\! 
          \gamma_\mu^\T \frac{\omega^4N_V}{\omega^4+k^4}
	  \frac{f_V}{M_V}\frac{Q^2}{Q^2+M_V^2}\, e^{-\alpha(Q^2+M_V^2)},
  \label{eqn:QEDVertexFitToLeadingTransverse}
\end{align}
which consists of the non-transverse Ball-Chiu part, $\Gamma^{\textrm{BC}}_\mu$, and the leading transverse structure
corresponding to $T_\mu^1=\gamma_\mu^\T$.
For the parameters we find reasonable agreement with the numerical solution with $\omega = 0.66\,\mbox{GeV}$, 
$\alpha=0.15$ and $N_V f_V/M_V = 0.152$.
Note that, as in the ENJL model, we have in Eq. (\ref{eqn:QEDVertexFitToLeadingTransverse})
a part that is given via the WTI (the BC vertex) and a transverse part.
We wish to emphasize once more, that the vector meson pole appearing in the 
transverse part is generated dynamically in the DSE and ENJL approaches.

\subsection{Differences between DSEs and ENJL model}

\begin{figure}[h]
  \begin{center}
    \includegraphics[width=0.3\textwidth,angle=-90]{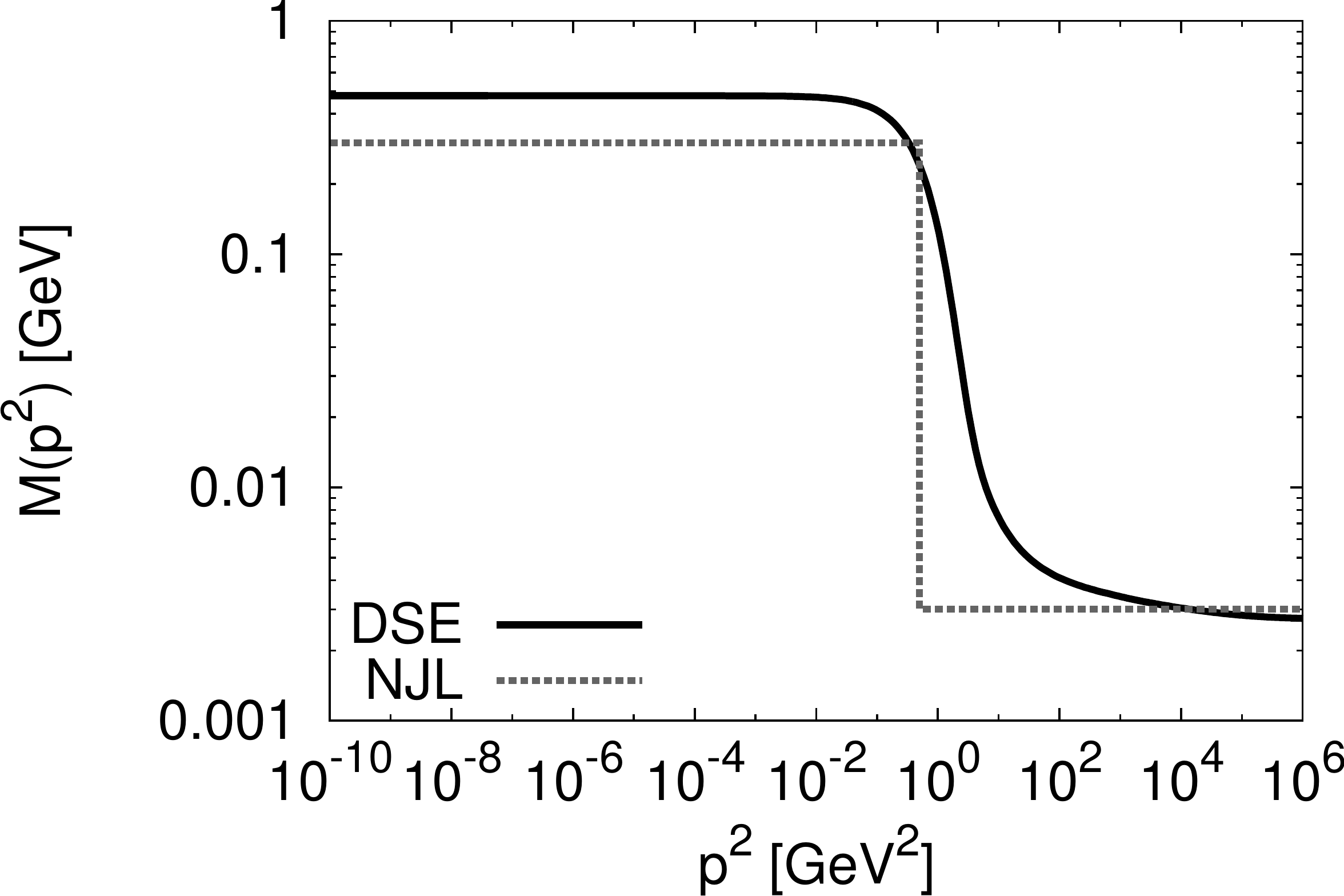}
    \includegraphics[width=0.3\textwidth,angle=-90]{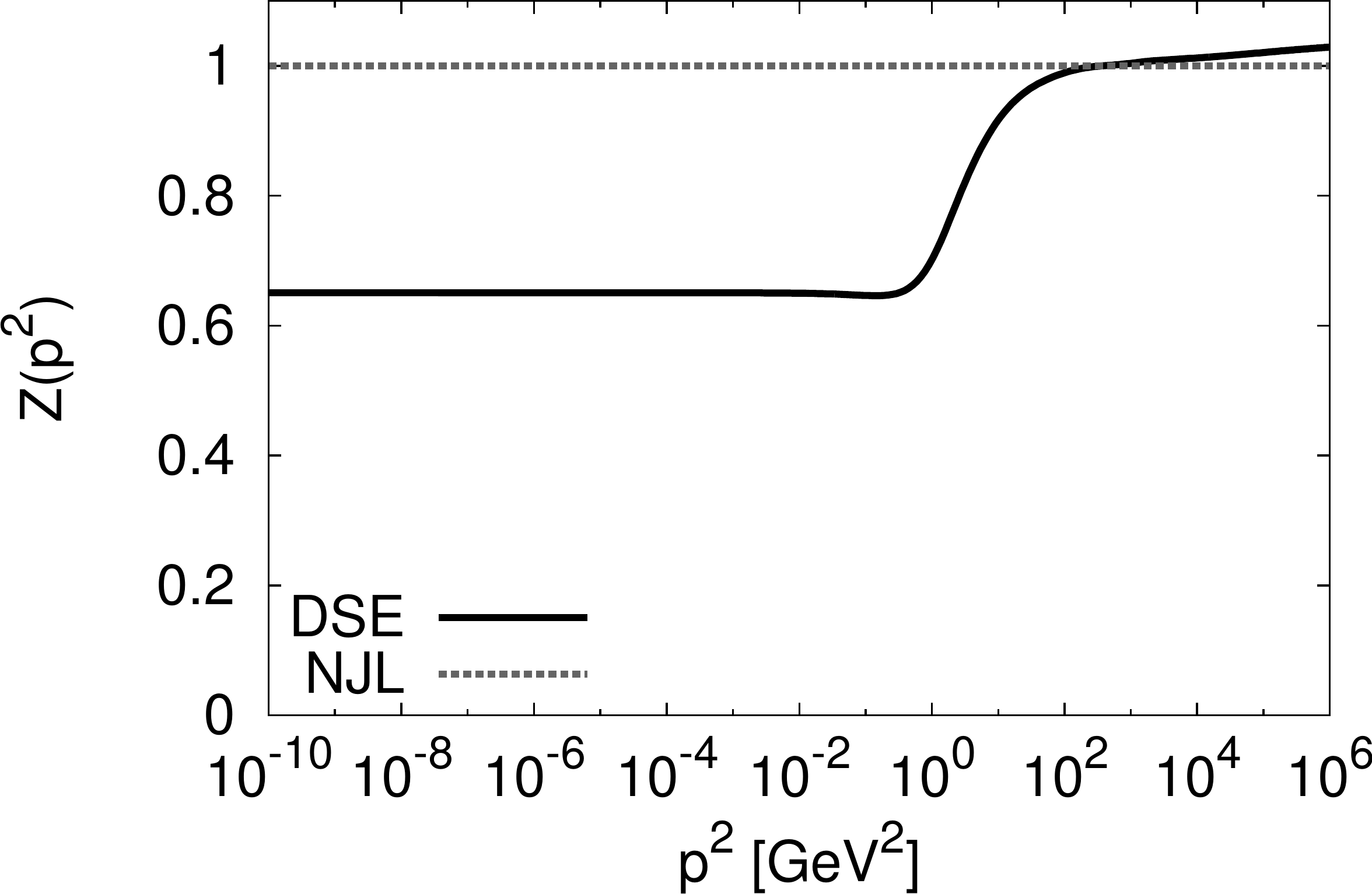}
  \end{center}
  \caption{Comparison of the mass function $M$ and the wavefunction
  $Z_f$ for DSE and ENJL quarks.}
  \label{fig:QuarkMandZdseANDnjl}
\end{figure}

We already stated that one of the chief 
differences between the DSE and ENJL approach lies with
the momentum dependencies of the propagators and vertices. We compare the
quark propagators in Fig.~\ref{fig:QuarkMandZdseANDnjl}, where one sees a
non-trivial momentum dependence in both the wave function renormalisation and
the mass function for the DSE. 
In particular, the mass function features 
the expected resumed one-loop perturbative behaviour at large momenta and 
smoothly connects to the infrared domain where a constituent-like mass 
is dynamically generated. This rapid accumulation of mass is consistent with
results of the operator product expansion~\cite{Fischer:2006ub}.  Indeed,
the same qualitative behaviour is seen on the lattice, see  Ref.~\cite{Fischer:2007ze}
for a comparison. In contrast, the dressing functions in the ENJL model are just
constants and hence miss essential features present in QCD.

\begin{figure}[h]
  \begin{center}
    \includegraphics[width=0.3\columnwidth,angle=-90]{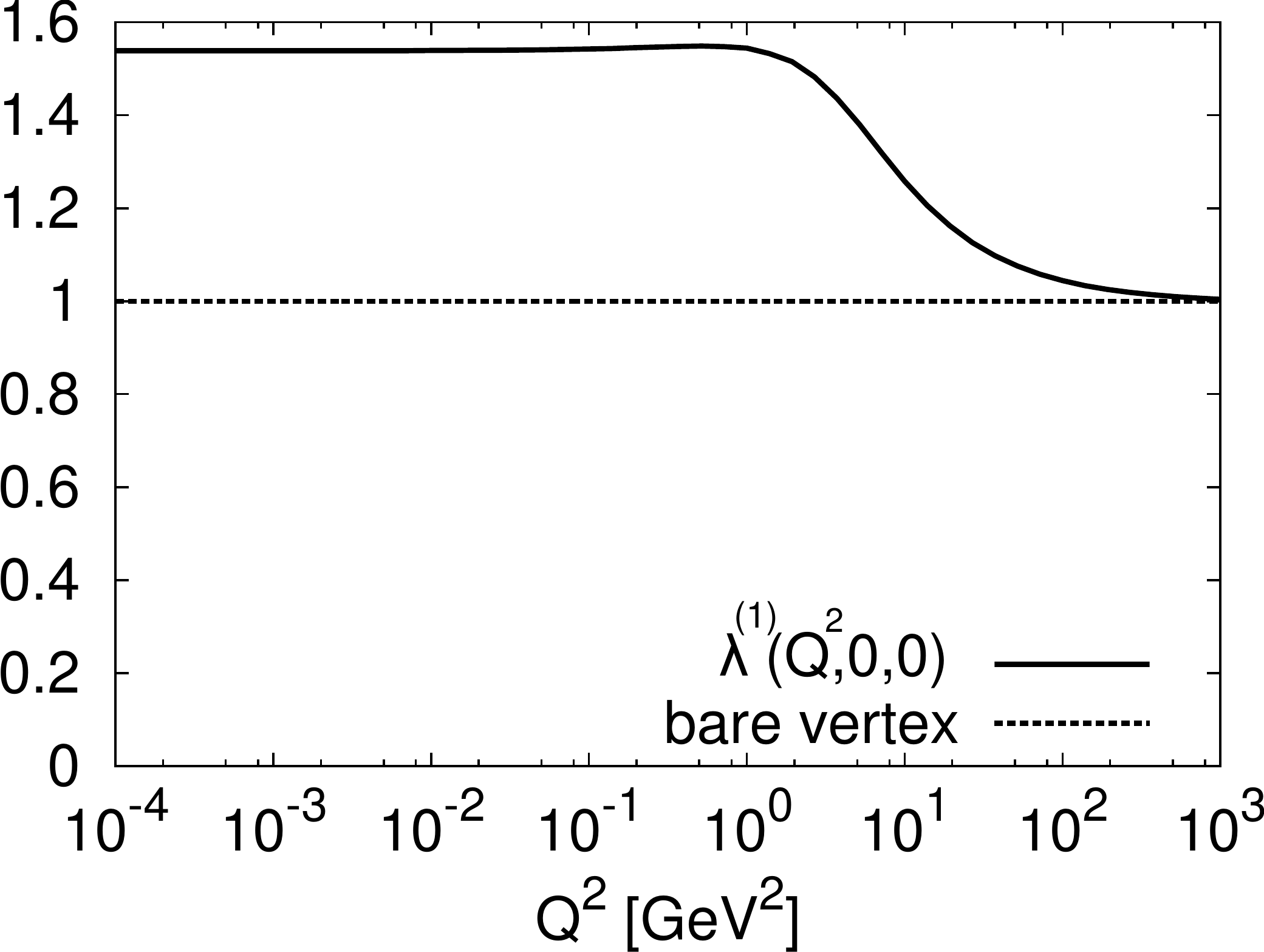}
    \includegraphics[width=0.3\columnwidth,angle=-90]{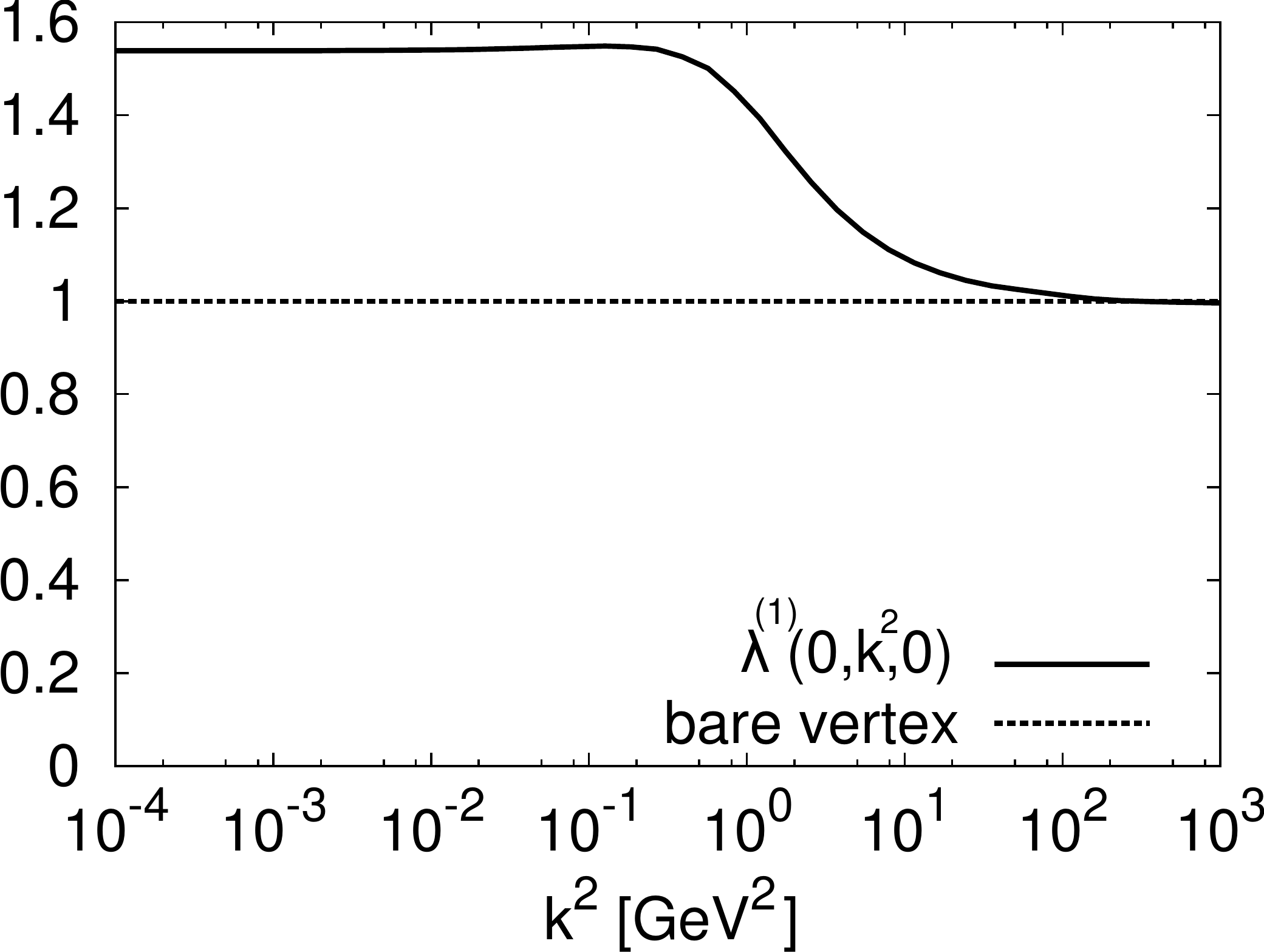}
  \end{center}
  \caption{The leading $\lambda^{(1)}$ component of the quark-photon vertex constrained
  by the Ward-Takahashi identity of  Eq.~(3.3).  We show two slices with relative, $k$, 
  and total, $Q$, momentum set to zero, respectively. The
  constant dressing corresponds to the ENJL model.}
  \label{fig:1BCVertexDSENJL}
\end{figure}

Next we compare the approximate functional forms of the quark-photon vertex
in the ENJL model, Eq.~(\ref{eqn:NJLVertexLT}) and DSE approach, Eq.~(\ref{eqn:VertexDecomposition}). The dressing of the $\gamma_\mu$ part is
shown in Fig.~\ref{fig:1BCVertexDSENJL}. Consistent with the Ward-Takahashi identity
it is unity in the ENJL case due to the quark wavefunction being trivial there. In
contrast, the corresponding vertex dressing in the DSE approach behaves as
$\lambda^{(1)}\sim 1/Z_f$, owing again to Ward identities.
A consequence of the wavefunction being screened for a large part of the momentum regime is
that $\lambda^{(1)}>1$ leading to an enhancement. However, since each vertex can be paired
with a quark propagator that features a screening $\sim Z_f$ these effects essentially
cancel in the quark-loop contribution to $a_\mu$. Thus, quantitative differences between
the ENJL model and the DSE approach owing to non-trivial wavefunction renormalisation and 
the $\lambda^{(1)}\gamma_\mu$ part
of the vertex are expected to be small. This is confirmed in explicit calculations below.

\begin{figure}[t]
  \begin{center}
    \includegraphics[width=0.35\textwidth,angle=-90]{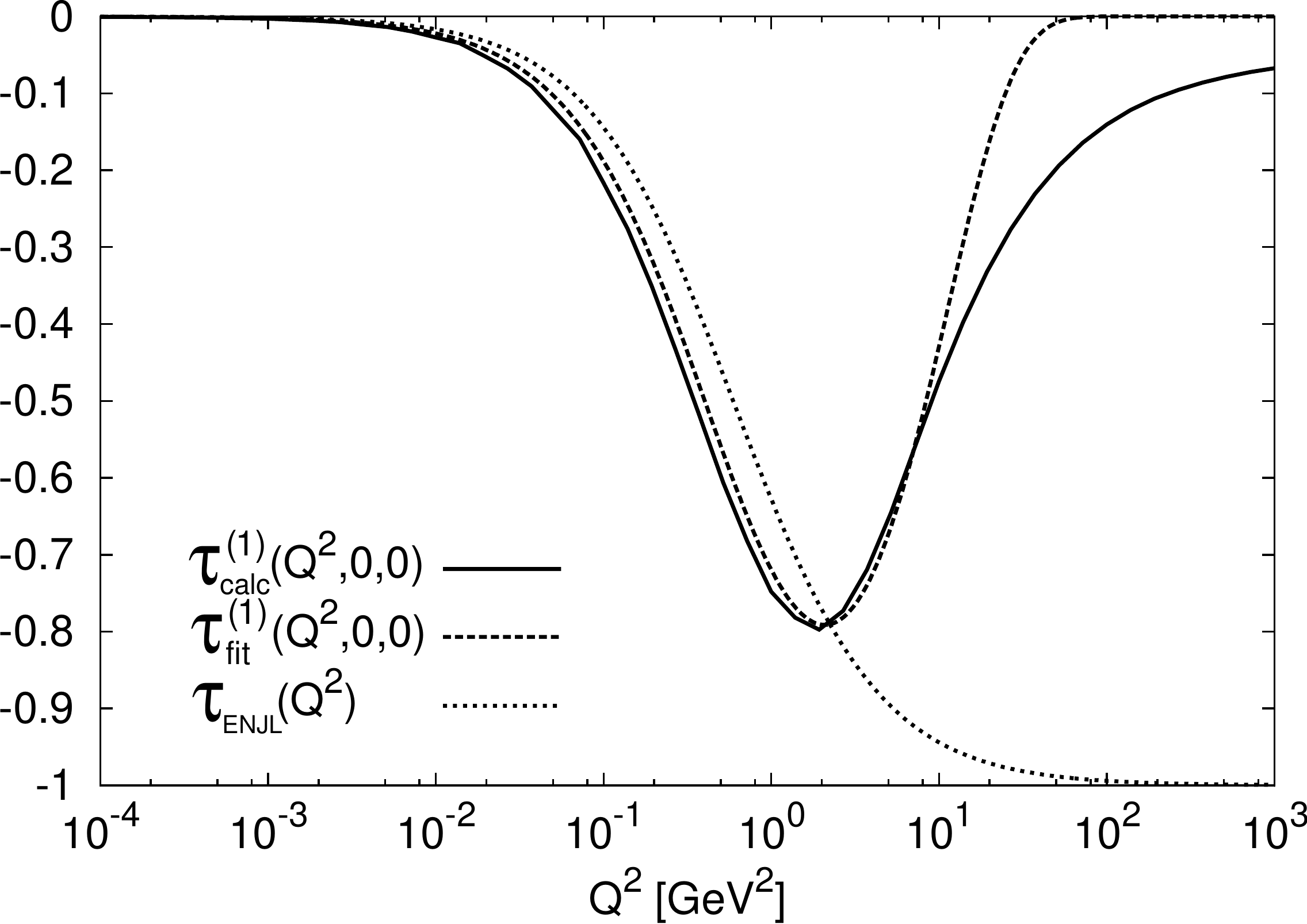}
  \end{center}
  \caption{The dependence of the dominant transverse dressing $\tau^{(1)}$ on 
  the space-like photon momentum $Q^2$ is shown for the explicit solution of the quark-photon
  vertex BSE, Eq. (3.2), the fit to this given in
  Eq.~(5.7), and the transverse part of the ENJL vertex,  
  Eq.~(5.3).}
  \label{fig:TransverseVertexDSENJL}
\end{figure}

Now we consider the transverse part of the vertex, $\tau^{(1)}=\gamma_\mu^\T$.
We compare the fit function of Eq.~(\ref{eqn:QEDVertexFitToLeadingTransverse}) with the full numerical solution,
$\tau^{(1)}_{\textrm{calc}}$, of the vertex BSE in Fig.~\ref{fig:TransverseVertexDSENJL}
for a momentum slice in which the relative momentum between the two quarks is \emph{vanishing}.
We see that at small photon momentum, $Q$, qualitative agreement between the ENJL and
DSE calculations. Since the ENJL model is not ultraviolet complete, it is not surprising
that large differences arise at large values of the photon momentum. Fortunately, under the
integral such a
momentum region is not weighted strongly in the determination of $a_\mu$ and so this
deviation is not important. This situation is drastically changed when one takes into account
the relative momentum between the two quarks, i.e. $k^2 \ne 0$. We see from the fit to the
DSE calculation of the vertex, Eq.~(\ref{eqn:QEDVertexFitToLeadingTransverse}) that the
transverse part of the vertex is highly suppressed for $k^2 > \omega^2$. The ENJL model
presents no such dependence. The quantitative impact of this damping will be discussed in
the next section.

\section{The hadronic light-by-light contribution}
\label{sec:results}

We are now ready to turn the qualitative discussion in the last section 
into quantitative statements. To this end we give a summary of the results of
Ref.~\cite{Goecke:2012qm} on the contribution to $a_\mu^{\textrm{LBL, quark-loop}}$
using various quark- and vertex-dressings. For some quantities we calculate
the averages of dressing functions as the arithmetic mean from samples of the
VEGAS Monte-Carlo routine \cite{Hahn:2004fe} we employ for $a_\mu^{\textrm{LBL, quark-loop}}$.
We consider the case of $N_f=2$ flavours only. The remaining flavours constitute 
a correction of the order of ten percent and don't change the general picture 
that emerges from this comparison.

\subsection{Influence of the quark and gauge part of the vertex}

Here, we explicitly show that the suspected cancellation between the
wavefunction renormalisation, $Z_f$, of the quark propagator and the leading gauge
part of the quark-photon vertex does, indeed, take place in the calculation of
the quark-loop contribution to $a_\mu$. Since the dressing
of the quark photon vertex is directly related via the Ward-Takahashi identity 
to $Z_f$ it suffices to perform the calculation in full, and with $Z_f=1/Z_2$\footnote{The $Z_2$ is a necessary renormalisation constant to ensure
multiplicative renormalizability}. We display the results in Table~\ref{tab:LBLResultsVertex}.
\begin{table}[b]
  \centering
  \begin{tabular}{l|c}
     													& $a_\mu^{LBL,ql} [10^{-11}]$	\\\hline
   $M$,\,$Z_f$,\,$\lambda^{(1)}$ all dynamical							&	$100$				\\
   $M$ dynamical,\,\,\,$Z_f=1/Z_2$,\,\,$\lambda^{(1)}=Z_2$ 					&	$102$				\\
   $M=0.2$ GeV,\, $Z_f=1/Z_2$,\,\,$\lambda^{(1)}=Z_2$ 					&     $104$		
    \end{tabular}
    \caption{The quark-loop contribution $a_\mu^{LBL,ql}$ to hadronic LBL. Compared are our result for full quark 
    propagator and gauge part of the vertex with an approximation where the momentum dependence of 
    the quark wave function is neglected. In addition we show the result where also the quark mass
    is replaced by a constant $M=0.2$ GeV.}
  \label{tab:LBLResultsVertex}
\end{table}
As is evident by the similarity of the two results, this clearly supports our
qualitative discussion regarding the cancellation.

In the process of performing these integrals, we also determined the average
mass probed to be $\sqrt{\langle M^2 \rangle} \approx 0.2$~GeV, as weighted by the Monte-Carlo integration. While this finding explains why constituent quark models
such as Ref.~\cite{Greynat:2012ww} and \cite{Masjuan:2012qn} must necessarily
use very light constituent quark masses in order to achieve reasonable results,
it also shows that this definition of effective mass is process dependent,
contrary to the running mass function defined through the quark DSE, which is universal.
The projection onto a certain process is only introduced through the process
dependent Feynman-diagrams
which includes the quark and thus the mass function.

We wish to emphasize that the $\lambda^{(1)}\gamma_\mu$ part is only one of three terms in
the gauge part of the vertex. Although the corresponding result is already approximately
gauge invariant \cite{Goecke:2012qm}, the functions $\lambda^{(2/3)}$ from Eq.~(\ref{eqn:VertexDecomposition})
have to be included for principal reasons. Unfortunately the corresponding numerics
turns out to be delicate \cite{Goecke:2010if}, such that more work is needed to
resolve this issue.

\subsection{Influence of the vertex: transverse part}

Now we investigate the impact of the leading transverse part of the vertex,
$\gamma_\mu^\T$, on the quark-loop contribution to $a_\mu$.

Since the ENJL model does not feature a relative momentum dependence in its
vertex dressings, we compare the ENJL result for the transverse vertex with
our DSE determinations where we force the condition $k^2=0$. From the 
left-panel in Table~\ref{tab:1BCandTransverseVertexResults} we see agreement
between the results which highlights that the differences at large momenta
in the fit functions are largely irrelevant.

In the right-panel of the same table, we show the corresponding results for
when the relative quark momentum is taken into account. An appropriate function is
also introduced to the ENJL parametrisation
\begin{align}
  f(k^2) = \frac{\omega^4}{k^4+\omega^4},
  \label{eqn:KsuppressionFunction}
\end{align}
with $\omega$ defined above. Again, we see a degree of parity between the 
models but note that there is more than a factor of two difference as compared
to the results with restricted momentum dependence.

\begin{table}[h]
  \centering
  \begin{tabular}{l|c}
	Vertex Dressing, $k^2=0$												&$a_\mu^{LBL,ql} [10^{-11}]$	\\\hline\hline
	$\gamma_\mu \lambda^{(1)}+\gamma^\T_\mu \tau_{\mathrm{ENJL}}$			&	$43$			\\\hline
	$\gamma_\mu \lambda^{(1)}+\gamma^\T_\mu \tau_\mathrm{fit}^{(1)}$		&	$43$			\\\hline
	$\gamma_\mu \lambda^{(1)}+\gamma^\T_\mu \tau_\mathrm{calc}^{(1)}$		&	$41$			\\\hline
  \end{tabular}
\hspace*{15mm}
  \centering
  \begin{tabular}{l|c}
	Vertex Dressing, $k^2 \ne 0$											&$a_\mu^{LBL,ql} [10^{-11}]$	\\\hline\hline
	$\gamma_\mu \lambda^{(1)}+\gamma^\T_\mu \tau_{\mathrm{ENJL}} f(k^2)$ 	&   $103$			\\\hline
	$\gamma_\mu \lambda^{(1)}+\gamma_\mu^\T \tau_\mathrm{fit}^{(1)} $		&	$105$			\\\hline
	$\gamma_\mu \lambda^{(1)}+\gamma^\T_\mu \tau_\mathrm{calc}^{(1)} $		&	$96$	
  \end{tabular}
    \caption{Leading gauge part and leading transverse vertex component, with dressing functions 
     from the ENJL model, VMD like fit from DSE/BSE ($\tau^{(1)}_{\textrm{fit}}$), and
     from an explicit calculation of the quark-photon DSE ($\tau^{(1)}_{\textrm{calc}}$).
     Results are shown without (left table), and with (right table) the inclusion of a 
     dependence on the relative momentum.}
  \label{tab:1BCandTransverseVertexResults}
\end{table}

Our most sophisticated vertex construction is the leading gauge part of the vertex 
plus the leading transverse part, corresponding to the last line 
of the right Table \ref{tab:1BCandTransverseVertexResults}. Repeating
this calculation for the case of $N_f=4$ flavours we obtain
\begin{align}
  a_\mu^{LBL,ql,N_f=4} = (107 \pm 2) \times 10^{-11}\,,
  \label{eqn:1BCplusLeadingTransverseLBLResultudsc}
\end{align}
where the error is purely statistical.

\section{Summary and conclusions}
\label{sec:conclusions}

We summarised the detailed comparison of the DSE framework to the ENJL model 
with respect to the quark loop part of the hadronic light-by-light contribution 
to the anomalous magnetic moment of the muon, $a_\mu^{LBL,ql}$, presented in 
Ref.~\cite{Goecke:2012qm}. We found similarities but also important differences. 
Our main focus was on the influence of the momentum dependence of dressing 
functions that are not present in the simplified treatment within the ENJL model. 
These are the quark mass and wave functions as well the leading non-transverse and 
transverse dressing functions of the quark-photon vertex. Since some of the
momentum dependencies are related via a Ward-Takahashi identity they cancel
approximately in the calculation of $a_\mu^{LBL,ql}$. Other important differences,
however, remain. One is the impact of the running quark mass function,
which turns out to be equivalent to an effective constituent quark mass
of $M\approx 0.2$ GeV, considerably 
smaller than typical values considered within
the ENJL model. This should, however, not be understood as a general
definition of a constituent mass, but rather a way how to make
a complicated integration simpler to get some intuitive insight into
the physics and numerics at work.
The other important difference concerns the impact of the
transverse parts of the vertex. In the ENJL model, the dependence of this
part of the vertex on the relative momentum $k$ between the quarks is neglected 
resulting in a large decrease of $a_\mu^{LBL,ql}$ due to transverse contributions.
Within the DSE approach we have shown that this suppression is dramatically 
reduced when the full momentum structure of the vertex is taken into account.

\begin{table}[b]
  \centering
  \begin{tabular}{l||c|c|c|c||c}
	$[10^{-11}]$						&$a_\mu^{LBL,(\pi_0)}$	&$a_\mu^{LBL,(\eta,\eta')}$	&$a_\mu^{LBL,ql}$	&$a_\mu^{LBL,other}$	&$a_\mu^{LBL}$ \\\hline\hline
	ENJL,Ref.\cite{Bijnens:1995xf}		&	\multicolumn{2}{c|}{85 (13)}					&       21 (3)		&		-23 (16)		&83 (32)	\\\hline
	standard,Ref.\cite{Prades:2009tw}	&	\multicolumn{3}{c|}{116 (13)}										&		-11 (13)		&105 (26)	\\\hline
	$\chi$QM,Ref.\cite{Greynat:2012ww}	&	68 					& - 						&		82 			&			-			&150 		\\\hline
	DSE, Ref.~\cite{Goecke:2012qm}		&	58 (1)				& 23 (1)					&		107 (2)		&			-			&188 (4)(90)\\\hline
  \end{tabular}
    \caption{Selected partial and total results for $a_\mu^{LBL}$. In the first line we display the ENJL result, the second line 
    displays the current 'standard result' (numerical and systematic error), the third line stems from a recent estimate 
    within a chiral quark model \cite{Greynat:2012ww} (evaluated at a constituent mass of $M=240$ MeV), whereas the fourth line shows
    our results so far (numerical error only, with a guess of the systematic error of the total result). Cells with a dash mark 
    contributions which have not (yet) been calculated in the corresponding approach. The column 'other' sums up contributions 
    from scalar and axialvector meson exchange as well as contributions from dressed pion loops.}
  \label{tab:finalresults}
\end{table}

In Table~\ref{tab:finalresults} we compare our current results for the different contributions to
$a_\mu^{LBL}$ with the ENJL-results of Ref.~\cite{Bijnens:1995xf}, the current `standard' result of 
Ref.~\cite{Prades:2009tw} (summarising calculations from different sources) and the values from a 
recent estimate within a chiral quark model \cite{Greynat:2012ww}, which has been advocated as a qualitative 
reference. Compared to the ENJL-model
we nicely agree in the meson exchange contributions $a_\mu^{LBL,(\pi_0)}$ and $a_\mu^{LBL,(\eta,\eta')}$
but note a drastic enhancement of the quark-loop contribution. We have explained this difference above. 
The same arguments apply to the results of the summary of Ref.~\cite{Prades:2009tw} in the second line.
The comparison with the chiral quark model ($\chi$QM) in the third line is interesting. They have applied 
a constituent mass in the range of $M=230-250$ MeV. As we have argued above, our results with full momentum 
dependence are reproduced by an approximation using an average quark mass function of 
$\langle M\rangle=200$ MeV, i.e. even lower.  For such a small value the results of Ref.\cite{Greynat:2012ww} 
become smaller in the pion exchange channel but larger in the quark-loop contribution resulting in a total 
of $168 \times 10^{-11}$ in very good agreement with our full results. 

Small constituent quark masses
of order $\sim 200$ MeV have been found to be necessary in various approaches using simple models, see e.g.
Refs. \cite{Pivovarov:2001mw,Boughezal:2011vw,Masjuan:2012qn}. As argued above, our findings can explain
this on a much deeper level of sophistication. It has to be mentioned, however, that not all models
agree on whether the quark-loop- and meson-exchange-contribution are two ways to describe the same contribution, or 
whether these have to be added. From the viewpoint of QCD diagrams it seems clear that these are completely
separate contributions, see e.g our Ref \cite{Goecke:2010if}.

We conclude from our study that the standard value $a_\mu^{LBL}= 105 (26) \times [10^{-11}]$ used in current evaluations of
the anomalous magnetic moment of the myon \cite{Hagiwara:2011af,Benayoun:2012wc} may be too small concerning its central value 
and is probably much too optimistic in its error estimate. We emphasize once again the importance of taking explicit momentum
dependencies into account, a task that the DSE framework is well capable for. Our current results for $a_\mu^{LBL}$ are preliminary,
since we neglected contributions from the transverse and non-transverse parts of the quark-photon vertex in the quark-loop.
We are working on this issue. We are also working on an explicit calculation of contributions from the charged pion-loop which,
according to the study of \cite{Engel:2012xb} may be more important than previously thought, see also the contribution of
Ramsey-Musolf to these proceedings \cite{RM}. 

To compete with the expected error bars of the forthcoming experiments we need to improve the theoretical error of $a_\mu^{LBL}$
into the twenty, or even better into the ten percent region. Within the DSE framework this level has been reached in the case of 
hadronic vacuum polarisation \cite{Goecke:2011pe}, where nice agreement with corresponding lattice calculations has been
obtained. For hadronic light-by-light we propose a similar strategy. We believe that only the systematic comparison of different 
methods like effective models, the DSE framework and lattice gauge theory provides the potential for a reliable calculation of $a_\mu^{LBL}$.    

{\bf Acknowledgments}\\
This work was supported by the DFG under contract FI 970/8-1
and the Austrian Science Fund FWF under project M1333-N16.

\end{document}